\documentclass[conference,a4paper]{APSIPA2019}
\usepackage{multirow}
\usepackage[dvips]{graphicx}
\usepackage{amsmath}
\usepackage[psamsfonts]{amssymb}
\usepackage{amsxtra}
\usepackage{threeparttable}
\usepackage{cite,graphicx,subfigure,amsmath,amsthm,amssymb,epsfig,pifont}
\usepackage{subfigure}
\usepackage{hyperref}
\usepackage{multirow}
\usepackage{epstopdf}
\usepackage{amsmath}
\usepackage{amsthm}
\usepackage{amsfonts}
\usepackage{float}
\begin{document}

\title{On Energy Compaction of 2D Saab Image Transforms}

\author{%
\authorblockN{%
Na Li\authorrefmark{1},
Yongfei Zhang\authorrefmark{2},
Yun Zhang\authorrefmark{1},
C.-C. Jay Kuo\authorrefmark{3}
}
\authorblockA{%
\authorrefmark{1}
Shenzhen Institutes of Advanced Technology, Chinese Academy of Sciences, Shenzhen, China \\
\authorrefmark{2}
School of Computer Science and Engineering, Beihang University, Beijing, China \\
\authorrefmark{3}
University of Southern California, Los Angeles, California, USA \\
}
}

\maketitle
\thispagestyle{empty}

\begin{abstract}

The block Discrete Cosine Transform (DCT) is commonly used in image and
video compression due to its good energy compaction property. The Saab
transform was recently proposed as an effective signal transform for
image understanding. In this work, we study the energy compaction
property of the Saab transform in the context of intra-coding of the
High Efficiency Video Coding (HEVC) standard.  We compare the energy
compaction property of the Saab transform, the DCT, and the
Karhunen-Loeve transform (KLT) by applying them to different sizes of
intra-predicted residual blocks in HEVC.  The basis functions of the Saab transform are visualized. Extensive experimental results are given to demonstrate the energy compaction capability of the Saab transform.

\end{abstract}

\section{Introduction}\label{sec:introduction}

Two-dimensional (2D) image transforms map 2D signals defined on a
regular grid from the spatial domain to the spectral domain. Since they
can compact the energy distribution in the spatial-domain to a fewer
number of frequency-domain coefficients, 2D image transforms are widely
used in image or video coding standards. Traditionally, 2D image
transforms adopt separable transform kernels.  That is, the 2D transform
kernel is formed by the tensor product of horizontal and vertical
one-dimensional (1D) transform kernels. The transforms are first
conducted along one dimension (say, the horizontal dimension) and then
followed in the second dimension (say, the vertical dimension).

It is well known that the Karhunen-Loeve transform (KLT) is the optimal
transform in the sense that it provides the best energy compaction
property.  To derive the 1D KLT, we first compute the covariance matrix
of image pixels in one horizontal or vertical segment of $N$ pixels
(say, $N=4$, $8$, $16$, $32$, {\em etc.}).  Next, we find the eigenvectors of
the covariance matrix, which are the KLT basis functions. The KLT is
also known as the principal component analysis (PCA).  The KLT is a
data-dependent (or data-driven) transform, which has higher
computational complexity than data-independent transforms. The discrete
cosine transform (DCT) \cite{Ahmed1974} is a data-independent transform,
which can be easily implemented by hardware for acceleration.  Besides,
the DCT provides a good approximation to data-dependent KLT for image
data.  The 2D separable DCT is widely used in today's image and
video coding standards.

We compare the energy compaction property of the Saab
transforms with the DCT and the KLT in the context of intra-coding of
the High Efficiency Video Coding (HEVC) standard \cite{Sull2012} in this
work. Our research objective is to enhance the DCT performance
by exploring three new directions.
\begin{enumerate}
\item Data-independent versus data-dependent transforms \\
Generally speaking, data-dependent transforms such as the PCA have better
energy compaction property than data-independent transforms such as the DCT. However,
data-dependent transforms received little attention in the past due to
their higher computational complexity. However, it is a recent trend in
modern video coding standards to trade higher computational complexity
for a higher coding gain. Thus, it is worthwhile to revisit data-dependent
transforms.
\item Separable versus nonseparable transforms \cite{Raj1978}, \cite{Aston2017} \\
Most existing higher dimensional transforms are derived as the tensor
product of several 1D transforms. However, this may not be the optimal
choice. It is actually easy to derive high dimensional transforms
directly. Let us use the 2D case as an example.  For an image block of
size $N \times N$, we can concatenate $N$ pixels in $N$ rows into one
long vector of length $N^2$ and compute the covariance matrix for such
random vectors accordingly. The eigenvectors of the covariance matrix
define a set of 2D nonseparable transform kernels. We would like to
check whether a 2D nonseparable transform is better than its corresponding 1D separable transform in energy compaction.
\item One-stage versus multi-stage transforms \\
By factorizing $N$ into the product of two integers, {\em i.e.}, $N=N_1 \times
N_2$, we can conduct two-stage PCA. The first stage PCA is applied to a
segment of $N_1$ pixels. There are $N_2$ segments in total. We obtain
$N_1$ transform coefficients for each segment. In the second stage, we
conduct the PCA on a data array of dimension $N_1 \times N_2$, where
$N_1$ and $N_2$ indicate the numbers of frequency and spatial
components. The output is still of dimension $N=N_1 \times N_2$.  One
example of multi-stage PCA is the Saab transform proposed by Kuo {\em et
al.} in \cite{Kuo2019}. The main feature of the Saab transform is to add
a sufficiently large constant to all output elements from the previous
transform stage so that all input elements to the next stage PCA are
non-negative.  This is needed to avoid the sign confusion problem. There
is little comparison study between one-stage and multi-stage PCA in the
literature \cite{Su2019}.  Here, we would like to see whether the Saab
transform ({\em i.e.}, multi-stage PCA) has a better energy compaction
property than one-stage PCA.
\end{enumerate}

We conduct extensive experimental results to demonstrate that the Saab
transform outperforms both the DCT and the PCA in terms of energy
compaction efficiency.  This makes it an attractive candidate in future
image/video coding standards.

The rest of this paper is organized as follows. Background is
reviewed in Sec. \ref{sec:background}. We examine the energy
compaction property of the DCT, the PCA and the Saab transform on different sizes of intra-predicted residual blocks in HEVC in Sec.
\ref{sec:compaction}.  The basis functions of the Saab transform are
visualized in Sec. \ref{sec:visualization}.  Concluding remarks are
given in Sec. \ref{sec:conclusion}.

\section{Background Review}\label{sec:background}

Transform energy compaction means the capability of a transform to
redistribute signal energy into a smaller number of transform
coefficients.  Among state-of-the-art separable
transforms, the DCT is the
most widely used in the compression field, including the JPEG image
coding standard and MPEG-1, MPEG-2, H.264/AVC and HEVC video coding
standards.  For example, in the latest HEVC coding standard, the DCT is
applied to both intra and inter block residuals of size $N \times N$,
where $N=4$, $8$, $16$ and $32$.  It is followed by quantization and
entropy coding. The kernel functions of the 2D DCT are in form of
\begin{equation}
\begin{array}{l}
f(m,n) = \\
\frac{2}{N} \sum\limits_{p = 0}^{N - 1}
\sum\limits_{q = 0}^{N - 1}\Lambda (p) \Lambda(q) {
\cos(\frac{(2p + 1)\pi m}{2N})
\cos(\frac{(2q + 1)\pi n}{2N})} ,
\end{array}
\end{equation}
where $m,n=0, \cdots, N-1$ and $\Lambda(\xi)= \sqrt{2}^{-1}$ if $\xi=0$
and $1$, otherwise.

To consider dependency of row and column elements of a picture
\cite{Raj1978}, one can arrange pixel samples in an image block of size
$N\times N$ in lexicographic order; namely,
\begin{equation}\label{eq:x}
\begin{array}{l}
{\textbf{x}} = [{x_{00}},\,\,{x_{01}},\,\,...\,,\,\,{x_{0,N - 1}},
{\kern 1pt} \,{x_{10}},\,\,{x_{11}},\,\,...,\,\,{x_{1,N - 1}},\\
\quad \quad \quad \quad \quad ...\,,\,\,{x_{N - 1,0}}\,,\,\,...\,,
{x_{N - 1,N - 1}}]^T.
\end{array}
\end{equation}
Similarly, we can express a transform kernel as
\begin{equation}\label{eq:a}
\begin{array}{l}
{\textbf{a}} = [{a_{00}},\,\,{a_{01}},\,\,...\,,\,\,{a_{0,N - 1}},
{\kern 1pt} \,{a_{10}},\,\,{a_{11}},\,\,...,\,\,{a_{1,N - 1}},\\
\quad \quad \quad \quad \quad ...\,,\,\,{a_{N - 1,0}}\,,\,\,...\,,
{a_{N - 1,N - 1}}]^T.
\end{array}
\end{equation}
One can conduct the PCA on random vectors as defined by Eq. (\ref{eq:x})
and choose the principal components as the kernels in Eq. (\ref{eq:a}).

Pixels in images have a decaying correlation property. The correlation
between local pixels is stronger and the correlation becomes weaker as
their distance becomes larger.  To exploit this property, Kuo {\em et
al.} \cite{Kuo2019} conducted a subspace affine transform in a local
window to get a local spectral vector. The input space is first
decomposed into the DC (direct current) subspace and the AC (alternate
current) subspace in the Saab transform. The AC subspace is formed by elements
\begin{equation}\label{eq:ACspace}
\begin{array}{l}
{\textbf{x}_{AC}}=\textbf{x}-({\textbf{a}}_0^T{{\textbf{x}}+{b_0}}){\bf 1},
\end{array}
\end{equation}
where ${\bf 1}={\bf c}/||{\bf c}||$, and ${\bf c}=(1, 1, \cdots, 1, 1)$
is the constant unit vector, in each stage of the Saab transform. The
block Saab transform is given by \cite{Kuo2019}
\begin{equation}\label{eq:saab}
\begin{array}{l}
{y_k} = \sum\limits_{n = 0}^{ N^2 - 1} {{a_{k,n}}{x_n} + {b_k}}  =
{\textbf{a}}_k^T{\textbf{x}} + {b_k}\quad k = 0,1,..., N^2 - 1 ,
\end{array}
\end{equation}
where ${{\textbf{a}}_0}$ is the DC filter and ${{\textbf{a}}_k},\;k = 1,..., N^2
- 1$ are the AC filters. The PCA is adopted to compute the AC filters in
the AC subspace. The bias term $b_k$, $k = 0,1,..., N^2 - 1$ is selected
to be a sufficiently large positive number to ensure that
${\textbf{x}_{AC}}$ is non-negative when it serves as the input to the
Saab transform in the next stage.

We would like to emphasize the main differences between the KLT and the Saab
transform below.
\begin{itemize}
\item The KLT does not decompose signals into DC and AC components. It
removes the ensemble mean and then compute the eigen-vectors of the
covariance matrix of mean-removed signals. The Saab transform has one
default DC filter. Then, we conduct the PCA on DC-removed signals.
\item The input to the KLT can be positive and negative values while the
input to any stage in the Saab transform should always be non-negative.
\end{itemize}

The Saab transform proposed by Kuo {\em et al.} \cite{Kuo2019} is a
data-driven (PCA-based), multi-stage, and nonseparable transform. It
meets all three criteria stated in Sec. \ref{sec:introduction}.  The
Saab transform is motivated by the analysis of nonlinear activation of
convolutional neural networks (CNNs) in \cite{Kuo2016,Kuo2017} as well
as the subspace approximation interpretation of convolutional filters in
\cite{Kuo2018}. We will focus on the energy compaction property of the
2D Saab image transform in this preliminary study and show that it
offers better energy compaction capability than the DCT and the KLT.
Thus, it is an attractive image transform candidate for image and video
compression.

To make our study as close to the real world coding environment as
possible, we study the energy compaction capability of several
transforms applied to block residuals obtained by intra prediction in
the HEVC test model version 16.9 under the all intra configuration in
the next section.

We consider non-overlapping blocks with the following settings.
\begin{itemize}
\item block size $4 \times 4$ \\
For the one-stage transform, we map 16 pixels in one block to one direct current (DC)
coefficient and 15 alternating current (AC) coefficients.  For the two-stage transform, we
first map one subblock of size $2 \times 2$ to one DC and 3 AC
coefficients.  Afterwards, we map a spatial-spectral cuboid, which has a
spatial dimension of $2\times 2$ and a spectral dimension of $4$, to a
spectral vector of dimension 16. Again, it has one DC and 15 AC
coefficients.
\item block size $8 \times 8$ \\
For the one-stage transform, we map 64 pixels in one block to one DC
coefficient and 63 AC coefficients.  For the two-stage transform, we
consider two cases. For the first case, we first map one subblock of
size $2 \times 2$ to one DC and 3 AC coefficients.  After that, we map a
spatial-spectral cuboid, which has a spatial dimension of $4\times 4$
and a spectral dimension of $4$, to a spectral vector of dimension 64.
For the second case, we first map one subblock of size $4 \times 4$ to
one DC and 15 AC coefficients.  Afterwards, we map a spatial-spectral
cuboid, which has a spatial dimension of $2\times 2$ and a spectral
dimension of $16$, to a spectral vector of dimension 64. In both of the
cases, it has one DC and 63 AC coefficients.
\item block size $16 \times 16$ \\
For the one-stage transform, we map 256 pixels in one block to one DC
coefficient and 255 AC coefficients.  For the two-stage transform, we
first map one subblock of size $4 \times 4$ to one DC and 15 AC
coefficients.  Afterwards, we map a spatial-spectral cuboid, which has a
spatial dimension of $4\times 4$ and a spectral dimension of $16$, to a
spectral vector of dimension 256, which contains one DC and 255 AC
coefficients.
\end{itemize}

\section{Energy Compaction Comparison of Image Transforms}\label{sec:compaction}

We compare the energy compaction properties of the DCT, the KLT and
the one-stage and two-stage Saab transforms in this section.  First, we
describe the energy compaction measure and elaborate on how to collect
block residuals from the HEVC encoder and obtain the Saab transform
kernels accordingly in Sec. \ref{subsec:setup}.  Then, we compare the
energy compaction property of different transforms applied to
intra-predicted block residuals of various sizes and resolutions in HEVC
in Sec. \ref{subsec:comparison}.

\subsection{Energy Compaction Measure and Sample Number Selection}\label{subsec:setup}

The Saab transform kernels are computed from the covariance matrix,
which will converge statistically as the number of samples increases.
We need to determine the number of samples required by the covariance
matrix computation.  A covariance matrix is said to converge if the
Frobenius-norm difference of two covariance matrices sampled by $M$ and
$M'$ samples ($M \approx M'$) becomes sufficiently small, {\em e.g.}, less
than $\epsilon = 1.5 \times 10^{-4}$.

\begin{figure}[!htb]
\centering
\includegraphics[width=0.45\textwidth]{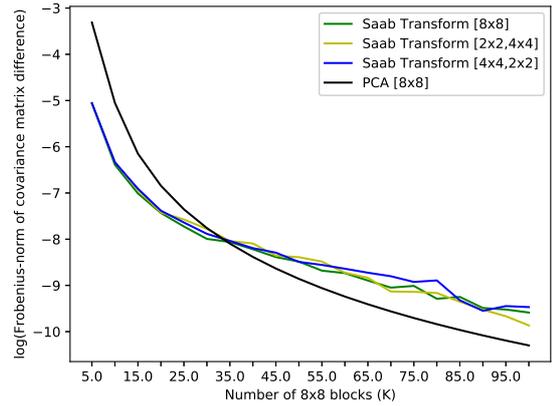}
\caption{Determination of the sample number required for Saab transform
kernel computation for the luminance (Y) block of size $8 \times 8$.}
\label{fig:convergency}
\end{figure}

An example of the relationship between the covariance matrix difference
computation and the number of samples is shown in Fig.
\ref{fig:convergency}, where the y-axis is the Frobenius-norm difference
of two covariance matrices expressed in the natural log scale and the
x-axis indicates the sample numbers.  This plot is obtained using the
$8\times8$ luminance (Y) residual blocks for sequence ``BasketballDrive"
of frame resolution $832\times480$, encoded by HM version 16.9 under all
intra configuration with four quantization parameter (QP) values ({\em
i.e.}, 22, 27, 32 and 37).  To compute the Saab transform kernels (or
the KLT eigenvectors), we see from the figure that it is sufficient to
use around 60K blocks for the covariance matrix to converge as
Frobenius-norm difference of two covariance matrices is to be less than
$\epsilon = 1.5 \times 10^{-4}$.  We adopt the same process in deriving
Saab transform kernels and KLT eigenvectors for block sizes with
different resolutions.

By energy compaction, we refer to the capability of a transform to
redistribute the signal energy into a small number of transform
coefficients. For a block of dimension $N \times N$, we obtain $N^2$
transform coefficients. For the DCT and the Saab transform, the first
one is the DC coefficient while the remaining ones are the AC
coefficients.  Although the KLT does not decompose signals into the DC
and AC subspaces, there is an equivalent concept; namely, its ensemble
mean component. The KLT subtracts the ensemble mean from all signals and
then compute the principal vectors
on the mean-removed signal subspace.  It is not
practical to compute the ensemble mean in image and video coding so that
we use the spatial mean to estimate the ensemble mean under the ergodic
assumption. By following the above condition, we show the averaged DC
energy ($k=0$), the AC energy ($1<=k<=15$) and the total energy ($0<=k<=15$) for 100 luminance (Y) block residuals of size $4\times4$ in Table
\ref{tab:T_4_Coef_Energy}, where the component $k=0$ in the KLT is the
energy of the ensemble mean vector.

\begin{table}[!htb]
\begin{center}
\caption{Averaged DC and AC energy values for luminance (Y) residual blocks
of size $4 \times 4$ under different transforms.}\label{tab:T_4_Coef_Energy}
\begin{tabular}{|c|c||c|c|c|c|} \hline
\multirow{2}{*}{Energy}&\multirow{2}{*}{Index (k)}& \multirow{2}{*}{DCT} &
\multirow{2}{*}{KLT} & \multicolumn{2}{c|}{Saab Transform} \\ \cline{5-6}
   & &  &  & [4$\times$ 4] & [2$\times$ 2, 2$\times$ 2] \\ \hline
DC & 0 & 122.35 & 150.29 & 122.35 & 49.53\\ \hline
AC & 1$\sim $15 & 155.56 & 127.02 & 155.19 & 228.20\\ \hline
Total & 0$\sim $15 & 277.90 & 277.31 & 277.54 & 277.73 \\ \hline
\end{tabular}
\end{center}
\end{table}

\begin{figure}[!htbp]
\centering
\includegraphics[width=0.4\textwidth]{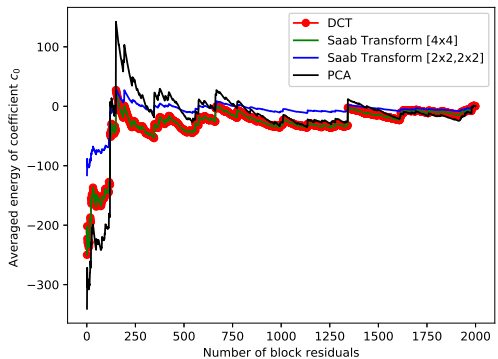}
\caption{Comparison of averaged energy of the coefficient indexed $k=0$ of DCT,
KLT, the one-stage and the two-stage Saab transforms as a function of the
number of residual blocks.} \label{fig:convergency_k0}
\end{figure}

Although there are small variations in the total energy values, the
difference becomes smaller as the sample size becomes larger.
Theoretically, the total energy before and after all orthonormal
transforms should be the same. Since the DCT and the one-stage Saab
transform compute the DC components of blocks in the same manner, they
have the same averaged DC values. The KLT has the highest energy for
$k=0$.  However, their differences become smaller as the number of
residual block samples increases as shown in Fig.
\ref{fig:convergency_k0}. Asymptotically, they all converge to zero since
we expect the residual blocks should be averaged out in the long run.

Interestingly, the two-stage Saab transform has the lowest DC energy
among the four in Table \ref{tab:T_4_Coef_Energy}. At the second-stage
Saab transform, its DC computation involves the sum of a cuboid of four
spatial dimensions and four spectral dimensions. Only the lowest
spectral dimension of the four spatial locations contribute significant
energy values.  Thus, after averaging, its DC value drops.  Since the DC
component has no discriminant power in image classification, the
multi-stage Saab transform is preferred for pattern recognition and
computer vision applications.  In the context of image and video coding,
DC and AC components are encoded separately. We expect the DC coding
cost is the lowest for the two-stage Saab transform.

When the number of block samples increase, the ensemble mean of the KLT
will converge to the DC value. As a result, the KLT, the DCT and the
one-stage Saab transform will have the same value for $k=0$. The
AC energy compaction is the only determining factor for the total energy
compaction. In contrast, the DC component of the two-stage Saab
transform is significantly smaller. The larger range of AC energy is good
for recognition. However, it is still not clear whether this helps or
hurts the overall RD gain. It demands further study.  If we show both DC
and AC in one plot, we cannot see the excellent AC energy compaction of
the two-stage Saab transform clearly since it is masked by its low DC
value. For this reason, we focus on the energy compaction property of a
transform of its AC coefficients only ({\em i.e.}, $K=1, \cdots, N^2-1$ for
$N^2$ coefficients) below.  Mathematically, we have
\begin{equation}\label{eq:energy_bands}
E_K^{N \times N} = \frac{{\sum\limits_{k = 1}^K c_k^2}}
{{\sum\limits_{k = 1}^{N^2-1} c_k^2}} \times 100\%,
\end{equation}
where $c_k$ is the coefficient of the $k$th AC component of the transform.

\begin{figure*}[htb]
\centering
\subfigure[``BasketballDrill"]{
\includegraphics[width=0.4\textwidth]{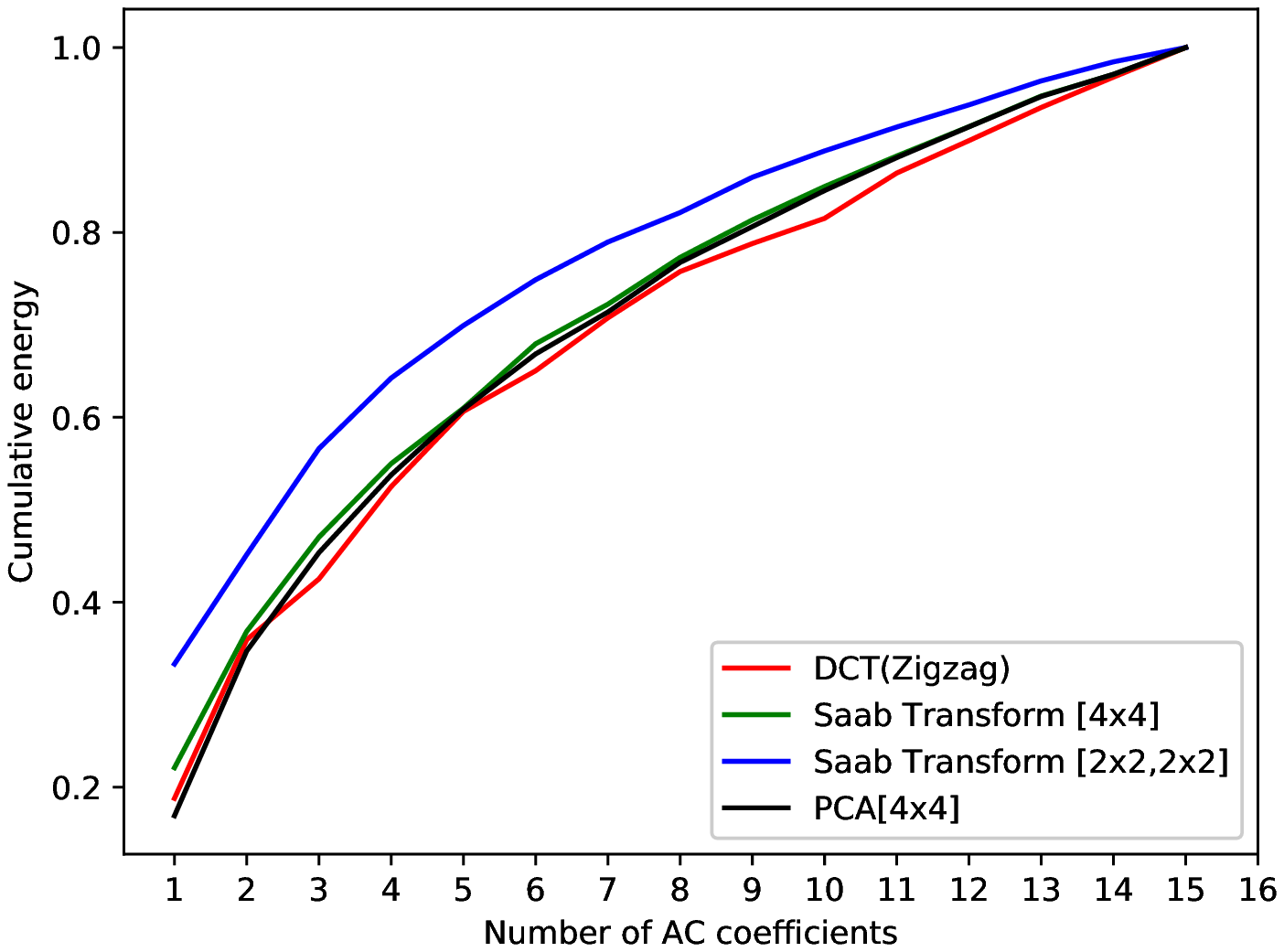}}
\subfigure[``FourPeople'']{
\includegraphics[width=0.4\textwidth]{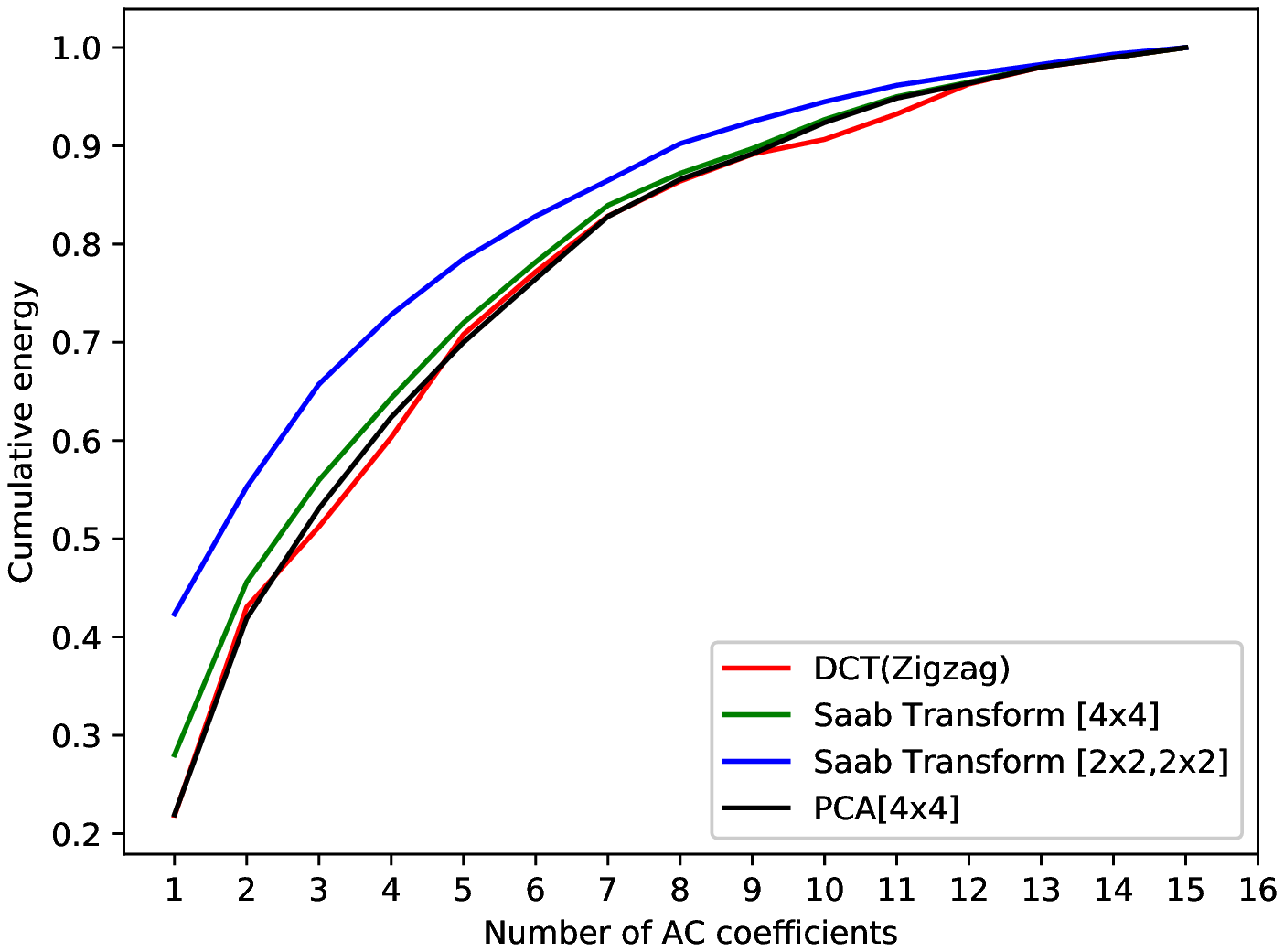}}
\subfigure[``BasketballDrive'']{
\includegraphics[width=0.4\textwidth]{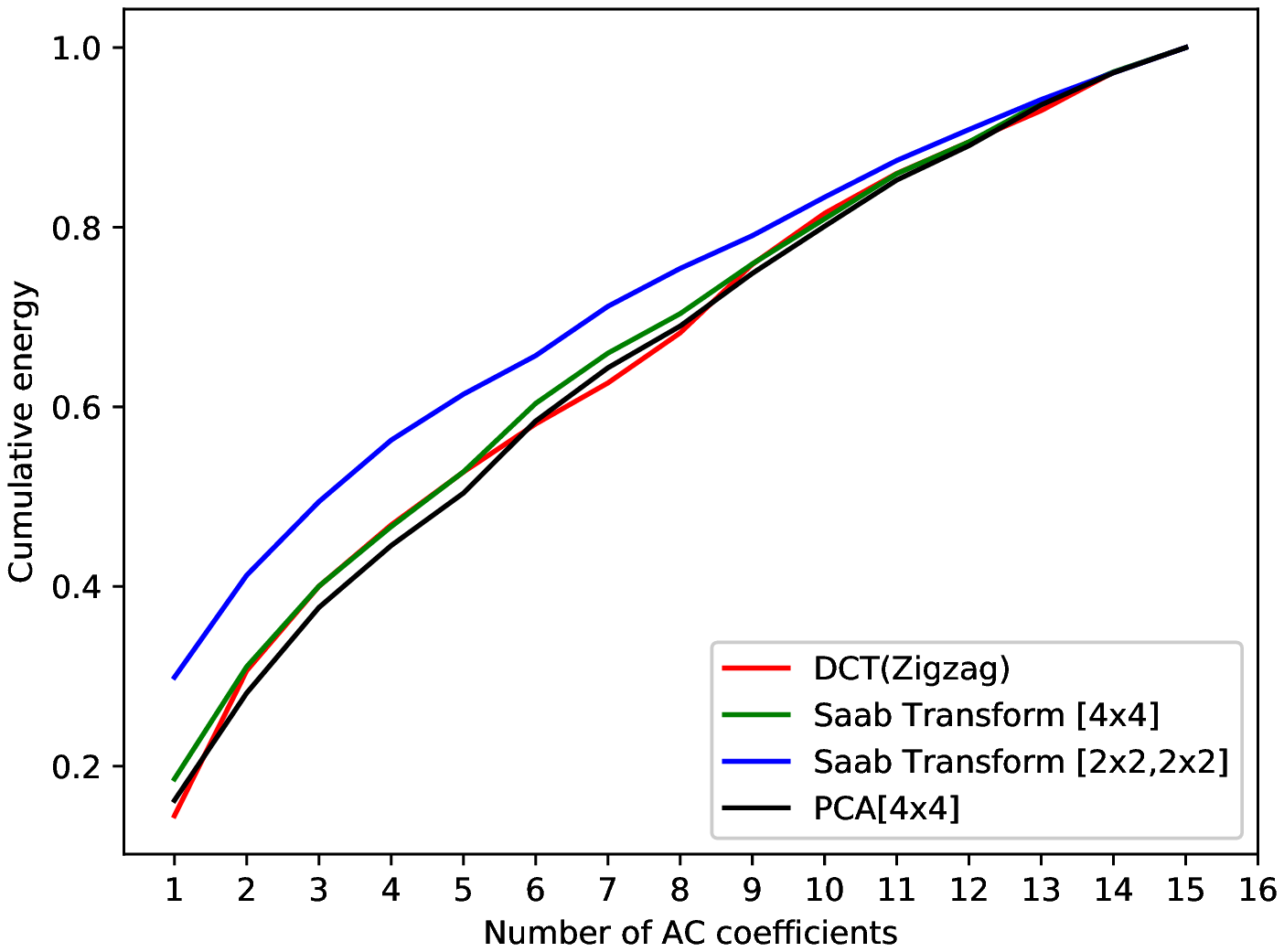}}
\subfigure[``PeopleOnStreet'']{
\includegraphics[width=0.4\textwidth]{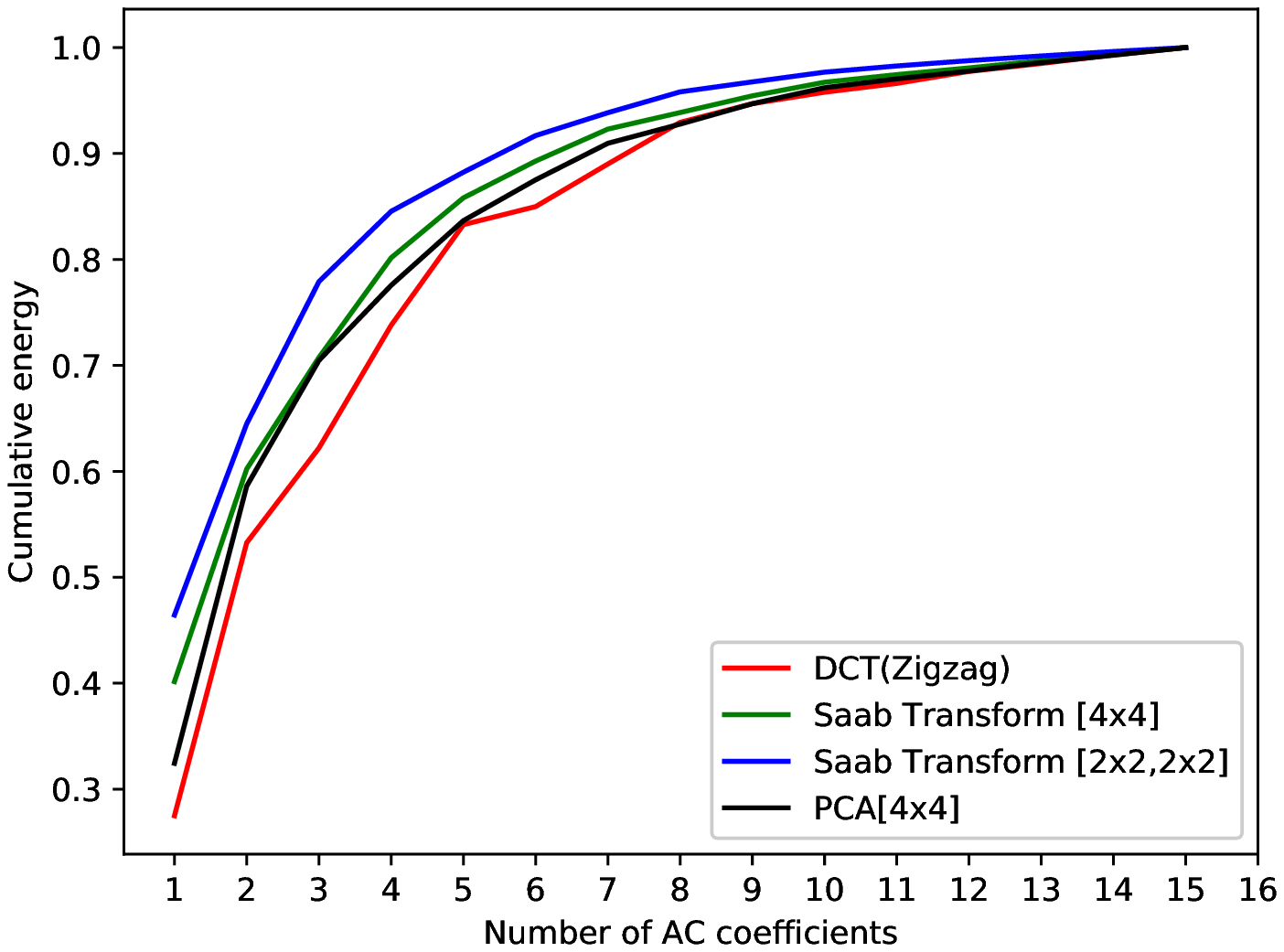}}
\caption{The cumulative AC energy plot for luminance (Y) block residuals
of size $4\times4$ for four video sequences: (a)``BasketballDrive" of
resolution $832\times480$, (b)``FourPeople" of resolution
$1280\times720$, (c)``BasketballDrive" of resolution $1920\times1080$,
and (d)``PeopleOnStreet" of resolution $2560\times1600$.}
\label{fig:Energy_Distribution_resolution_4Y}
\end{figure*}

\begin{figure*}[htb]
\centering
\subfigure[``BasketballDrill"]{
\includegraphics[width=0.4\textwidth]{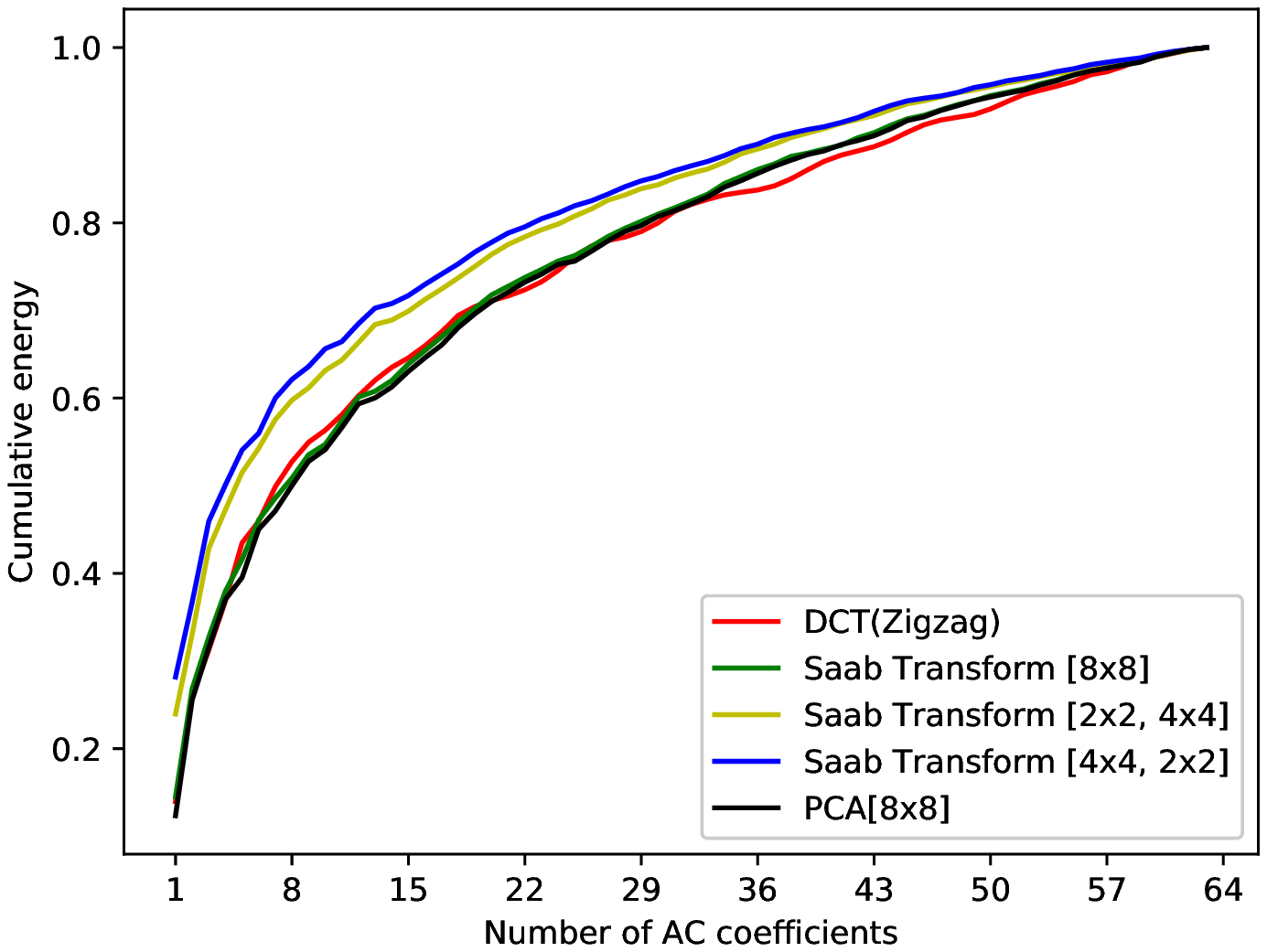}}
\subfigure[``FourPeople"]{
\includegraphics[width=0.4\textwidth]{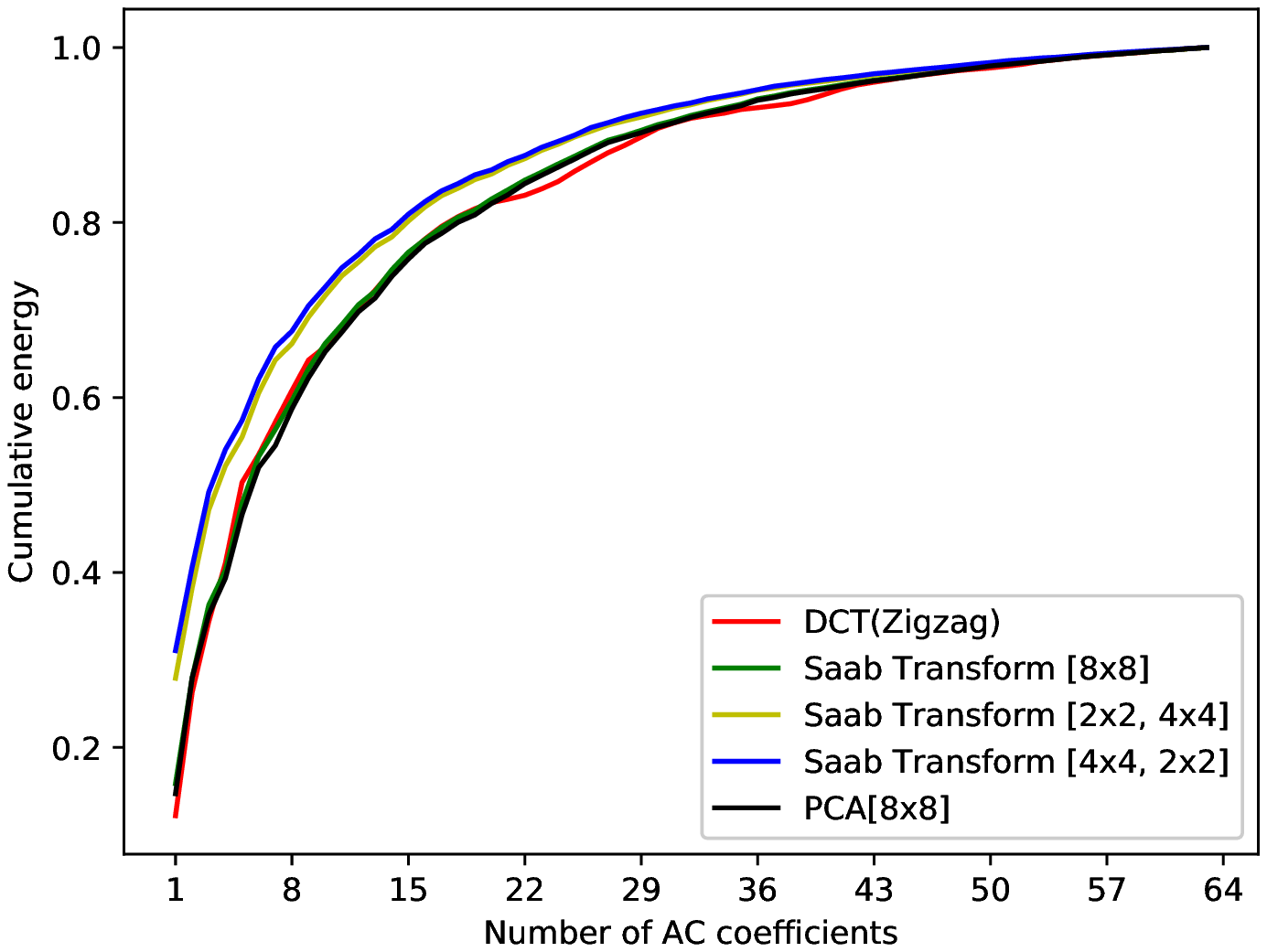}}
\subfigure[``BasketballDrive"]{
\includegraphics[width=0.4\textwidth]{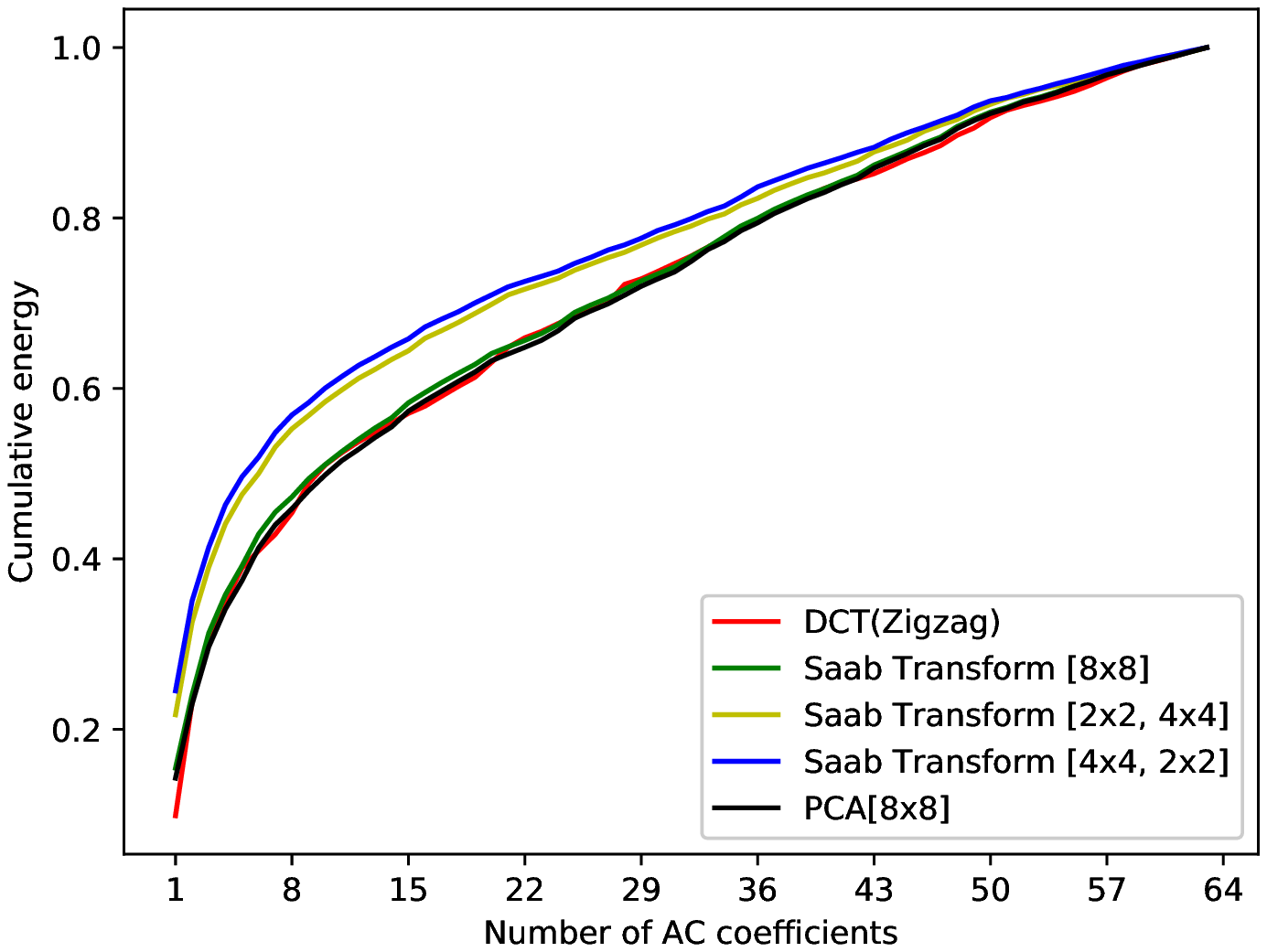}}
\subfigure[``PeopleOnStreet"]{
\includegraphics[width=0.4\textwidth]{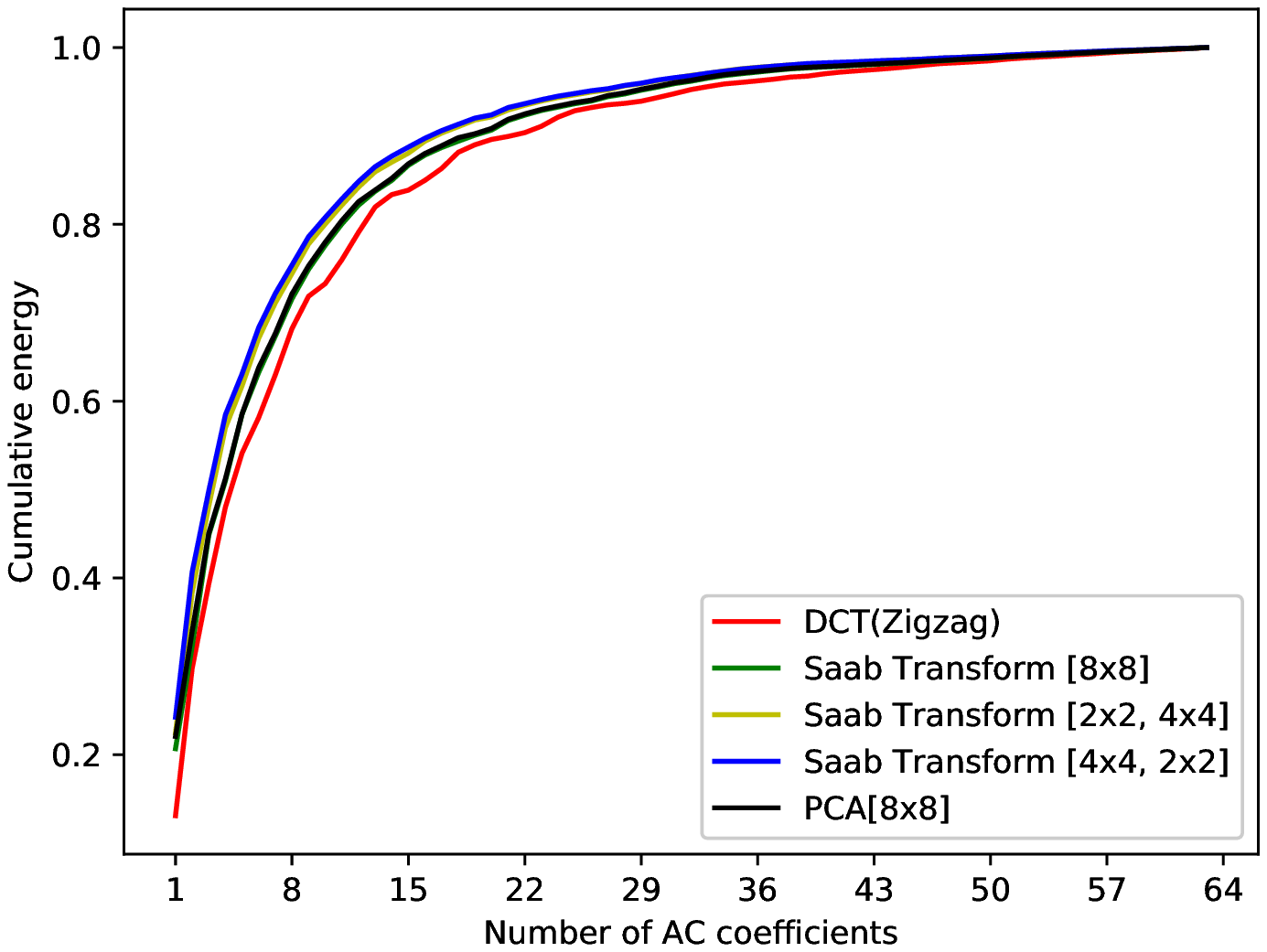}}
\caption{The cumulative AC energy plot for luminance (Y) block residuals
of size $8\times8$ for video sequences: (a)``BasketballDrill" of
resolution $832\times480$, (b)``FourPeople" of resolution
$1280\times720$, (c)``BasketballDrive" of resolution $1920\times1080$,
and (d)``PeopleOnStreet" of resolution $2560\times1600$.}
\label{fig:Energy_Distribution_resolution_8Y}
\end{figure*}

\subsection{Energy Compaction Performance Comparison}\label{subsec:comparison}

We compare the AC energy compaction performance of the DCT, the KLT,
one-stage and two-stage Saab transforms.  We will discuss the luminance
(Y) block residuals first and, then, the chrominance red (Cr) block
residuals.

\noindent
{\bf Luminance Blocks of Size $4 \times 4$.} We first examine residuals
for blocks of size $4 \times 4$.  They are obtained via intra prediction
coded with $QP=22$ and predicted as the planar mode. The cumulative AC
energy is plotted as a function of first $K$ AC coefficients in Fig.
\ref{fig:Energy_Distribution_resolution_4Y} (a). The curves indicate the
mean values of four different transforms.  We see that the DCT, the KLT
and the one-stage Saab transform have similar AC energy compaction
property while the two-stage Saab transform outperforms all of them by a
significant margin. The advantage of two-stage Saab transform on the
cumulative AC energy is demonstrated via sequences ``FourPeople",
``BasketballDrive" and ``PeopleOnStreet" of resolutions $1280\times720$,
$1920\times1080$ and $2560\times1600$ in Figs.
\ref{fig:Energy_Distribution_resolution_4Y} (b)-(d).

\noindent
{\bf Luminance Blocks of Size $8 \times 8$.} We examine luminance (Y)
block residuals of size $8\times8$ for four video sequences
with the intra predicted planar mode; namely,
``BasketballDrill", ``FourPeople", ``BasketballDrive" and
``PeopleOnStreet". The AC cumulative energy plots in Figs.
\ref{fig:Energy_Distribution_resolution_8Y}(a)-(d).  We have two
observations. First, the two-stage Saab transforms have better AC energy
compaction property than the DCT, the KLT and the one-stage Saab
transform. Two cases of the two-stage Saab transform are compared: 1)
$2\times 2$ spatial blocks followed by $4\times 4$ spatial blocks and 2)
$4\times 4$ spatial blocks followed by $2\times 2$ spatial blocks. Case
(2) is slightly better than case (1).

\begin{figure*}[htb]
\centering
\subfigure[``BasketballDrill"]{
\includegraphics[width=0.4\textwidth]{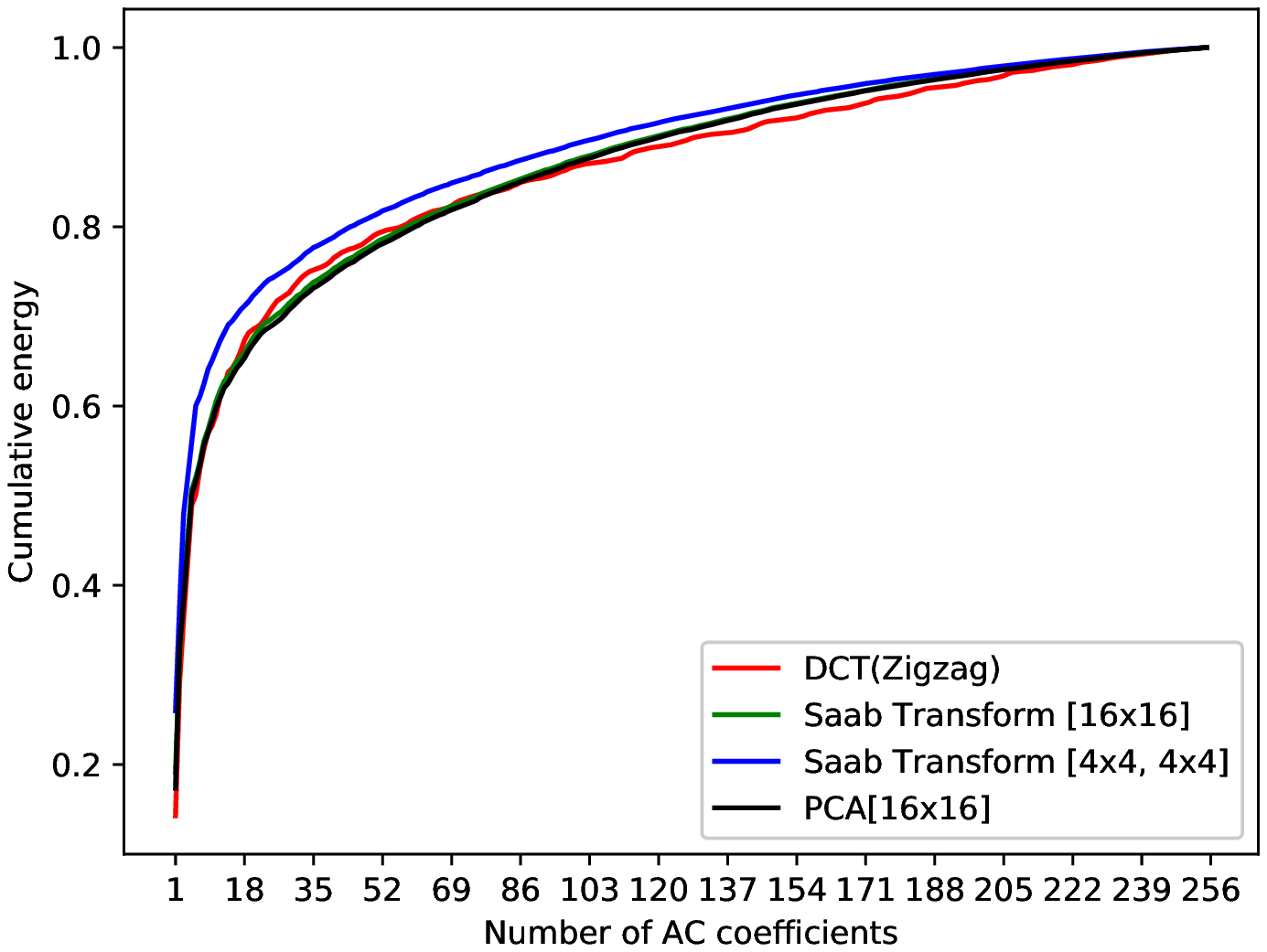}}
\subfigure[``FourPeople"]{
\includegraphics[width=0.4\textwidth]{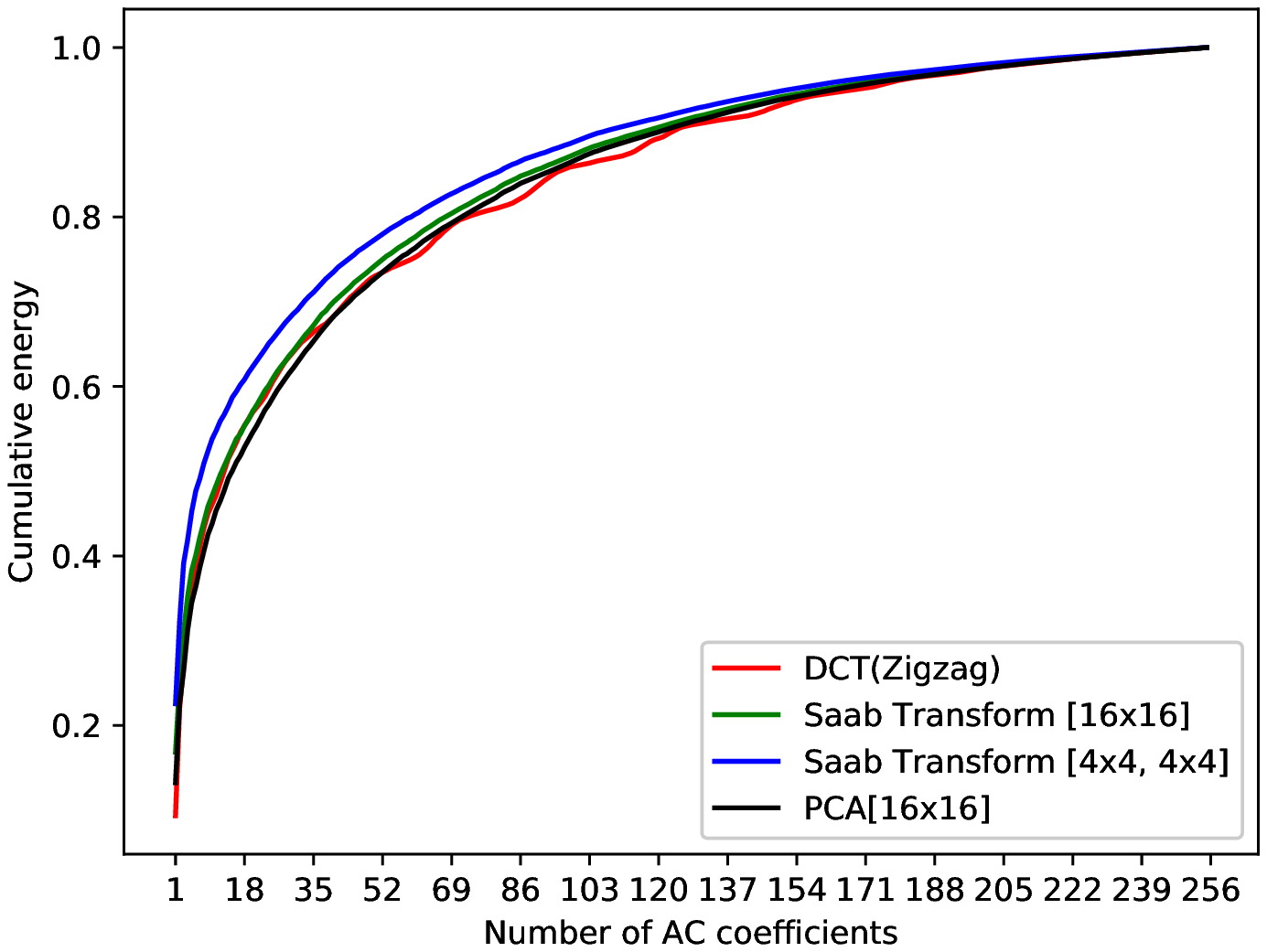}}
\subfigure[``BasketballDrive"]{
\includegraphics[width=0.4\textwidth]{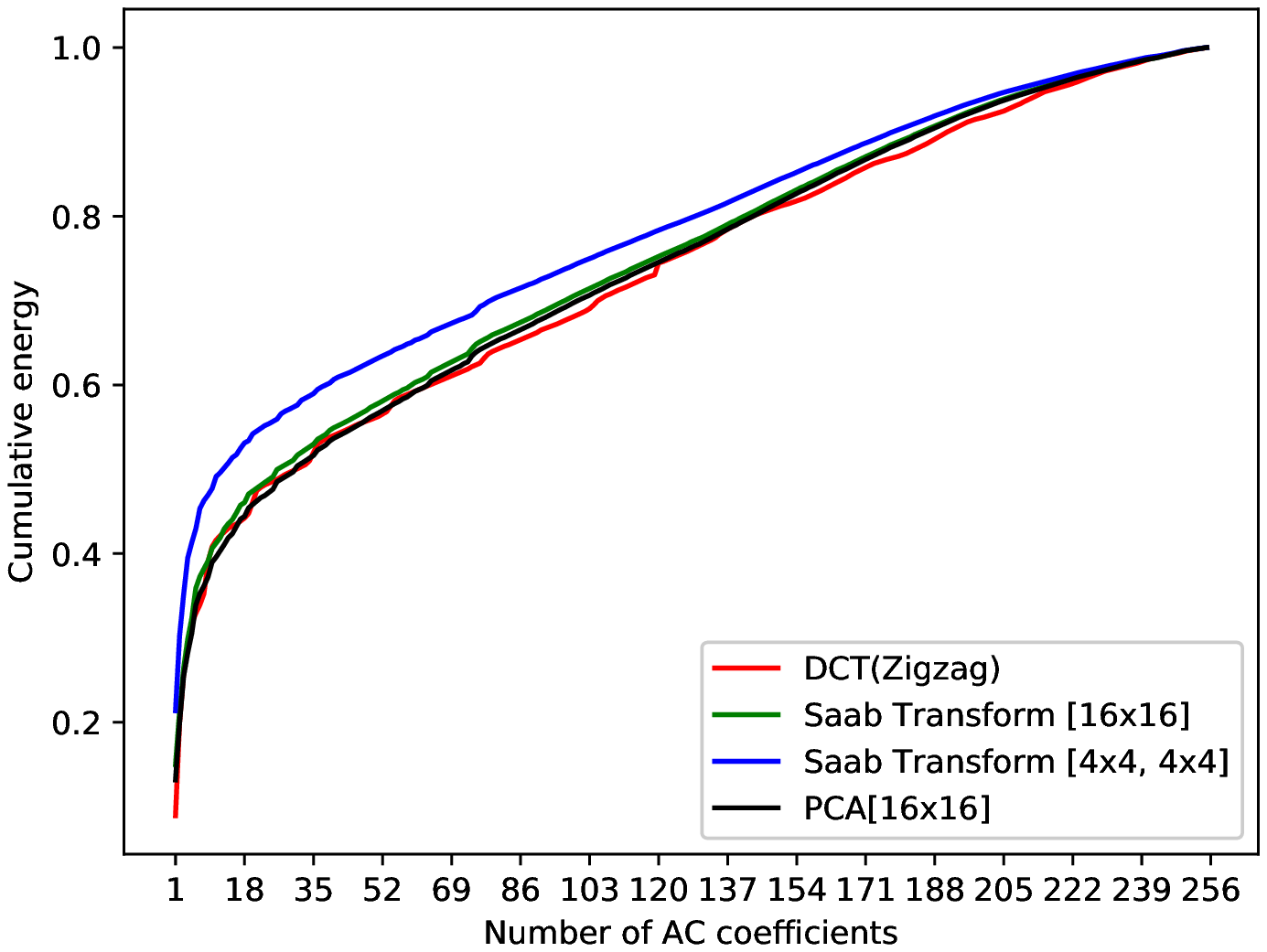}}
\subfigure[``PeopleOnStreet"]{
\includegraphics[width=0.4\textwidth]{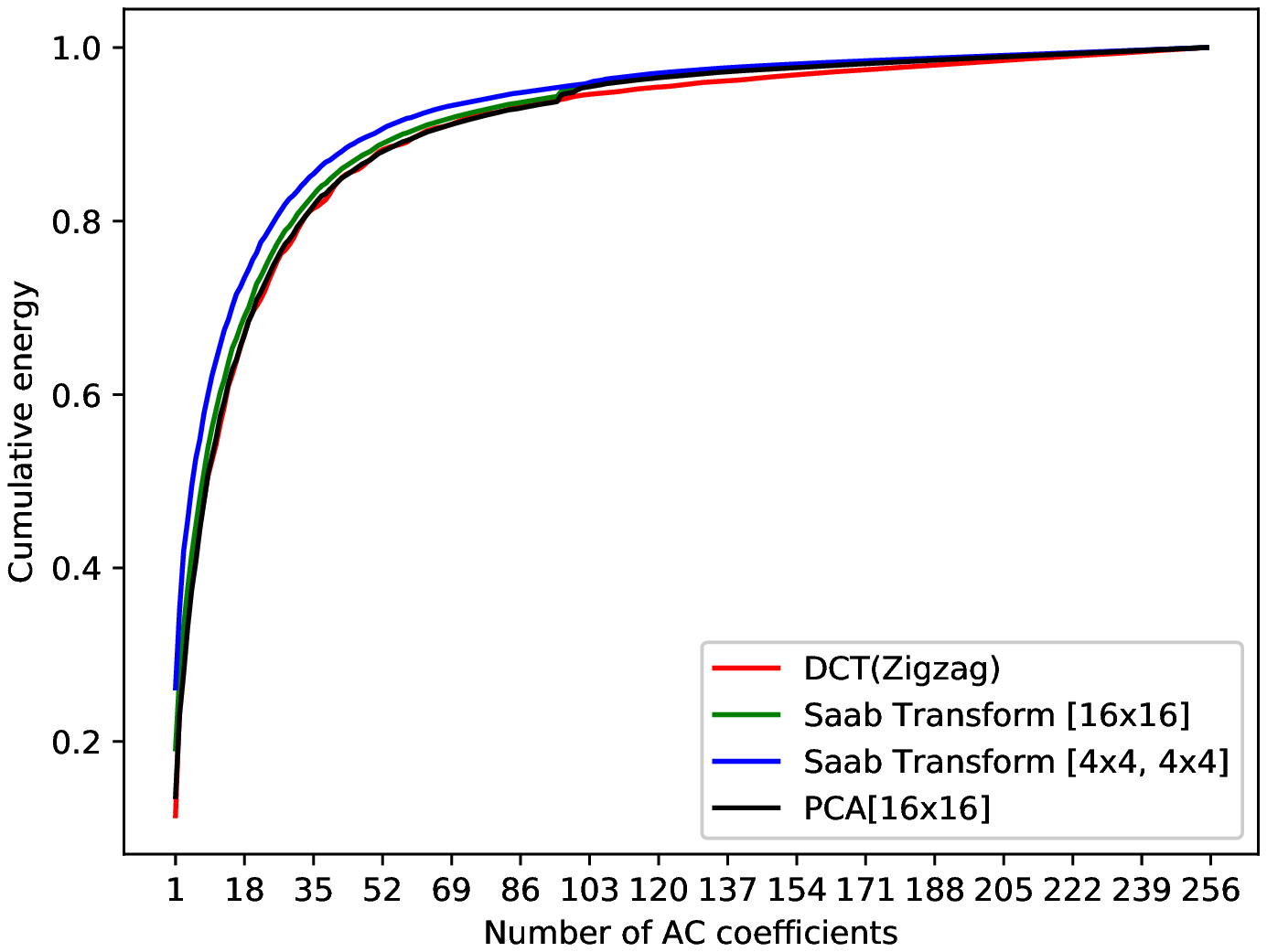}}
\caption{The cumulative AC energy plot for luminance (Y) block residuals of
size $16\times16$ for four video sequences:
(a)``BasketballDrill" of resolution $832\times480$,
(b)``FourPeople" of resolution $1280\times720$,
(c)``BasketballDrive" of resolution $1920\times1080$, and
(d)``PeopleOnStreet" of resolution $2560\times1600$.}
\label{fig:Energy_Distribution_resolution_16Y}
\end{figure*}

\noindent
{\bf Luminance Blocks of Size $16 \times 16$.} We plot the AC energy
compaction of luminance (Y) block residuals of size $16\times16$ with
the intra predicted planar mode for four video sequences of different
resolutions in Figs. \ref{fig:Energy_Distribution_resolution_16Y}(a)-(d).
The two-stage Saab transform with $4 \times 4$ spatial blocks followed
by $4\times 4$ spatial blocks has the best AC energy compaction
property among all benchmarking cases.

\begin{figure*}[!htb]
\centering
\subfigure[$8\times8$ chrominance red (Cr) block ]
{\includegraphics[width=0.4\textwidth]{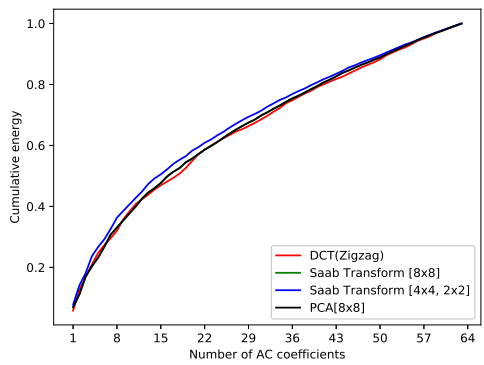}}
\subfigure[$16\times16$ chrominance red (Cr) block]
{\includegraphics[width=0.4\textwidth]{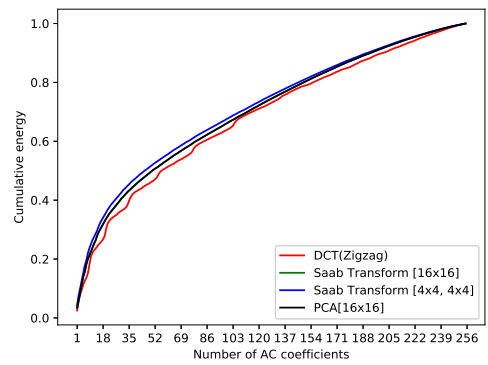}}
\caption{The cumulative AC energy plot for block Cr residuals under the
intra prediction mode for ``BasketballDrive" of resolution
$1920\times1080$ with block size: (a) $8\times8$ and (b) $16\times16$.}
\label{fig:Energy_Distribution_resolution_8Cr_16Cr}
\end{figure*}

\noindent
{\bf Chrominance Blocks of Sizes $8 \times 8$ and $16 \times 16$.} The
AC energy compaction properties for residuals of intra predicted
chrominance red (Cr) blocks for sequence ``BasketballDrive"  are compared
in Figs. \ref{fig:Energy_Distribution_resolution_8Cr_16Cr}(a)(b).
The differences between different transforms are smaller for chrominance red (Cr) block residuals.

\section{Visualization of Transform Basis Functions}\label{sec:visualization}

We can gain additional insights into image transforms by
visualizing their basis functions (or transform kernels). To obtain the
one-stage or the two-stage Saab transform basis functions, we set one
and only one AC spectral component in the last stage to unity while
setting other AC spectral components to zero. Afterwards, we perform the
inverse one-stage or two-stage Saab transform from the spectral domain
back to the spatial domain.  Finally, we normalize the gray level of
each pixel to the range between 0 and 255 using a linear scaling
operation followed by a shifting operation.

\begin{figure*}[!htb]
\centering
\subfigure[DCT]
{\includegraphics[width=0.35\textwidth]{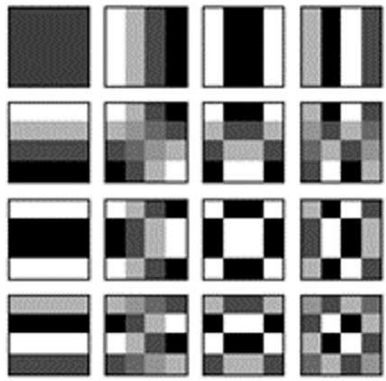}}
\subfigure[one-stage Saab transform]
{\includegraphics[width=0.45\textwidth]{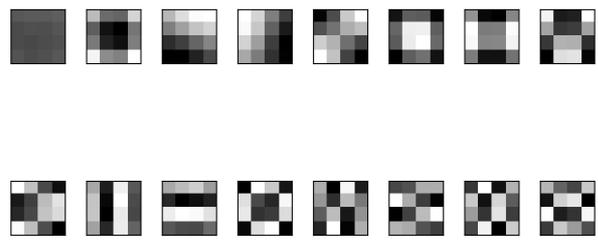}}
\subfigure[two-stage Saab transform]
{\includegraphics[width=0.45\textwidth]{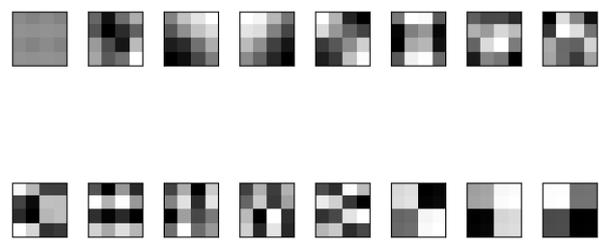}}
\caption{Visualization of transform basis functions for $4\times4$ blocks: (a)
the DCT, (b) the one-stage Saab transform, and (c) the two-stage Saab transform.}
\label{fig:44Basis}
\end{figure*}

\noindent {\bf Luminance Blocks of Size $4 \times 4$.} We compare the
basis functions of the DCT, the one-stage Saab transform and the
two-stage Saab transform with the planar mode in Fig.
\ref{fig:44Basis}. Since the DCT is a separable transform, we show the
transform basis functions in form of a 2D array in Fig.
\ref{fig:44Basis}(a), where the upper left corner is the DC component
while the other 15 are AC components. We see that separability of
transform kernels imposes severe constraints on the kernel form.  Since
the Saab transform is a nonseparable one, their transform kernels for DC
is followed by AC kernels, where AC kernels are ordered from the largest
to the smallest energy percentages, in two rows ({\em i.e.}, from left to
right and then from top to bottom). The basis function of the one-stage
Saab transform are shown in Fig.  \ref{fig:44Basis}(b). The first AC
basis is in bowl form, the second and the third AC basis functions are
tilted planes of 135 and 45 degrees, respectively, the fourth one is in
saddle form and the fifth one is in well form. These patterns are more
likely to happen. The remaining basis functions are less frequently
seen. Finally, we show the basis function of the two-stage Saab
transform Fig. \ref{fig:44Basis}(c).  The first AC of the two-stage Saab
transform is in bowl form but rotated by 45 degree. The most interesting
observation is the last three AC components.  The two-stage Saab
transform can include them in the basis function set. This is the major
difference between the one-stage and the two-stage Saab transforms.

\begin{figure*}[!htb]
\centering
\subfigure[intra horizontal]
{\includegraphics[width=0.8\textwidth]{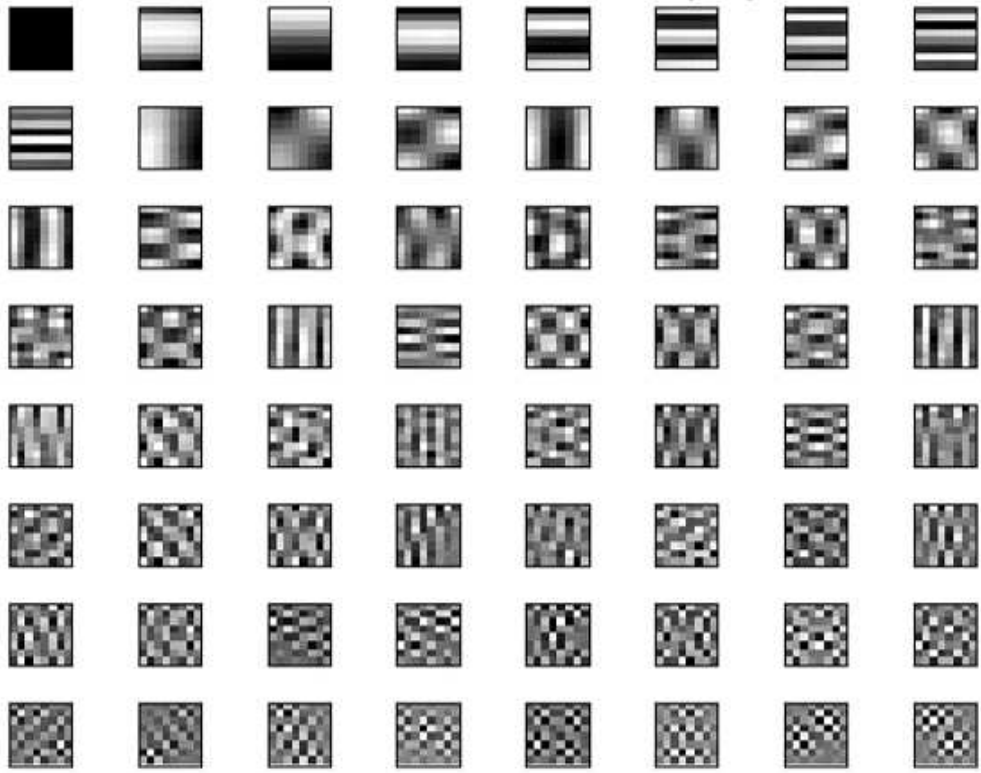}}
\subfigure[intra planar]
{\includegraphics[width=0.8\textwidth]{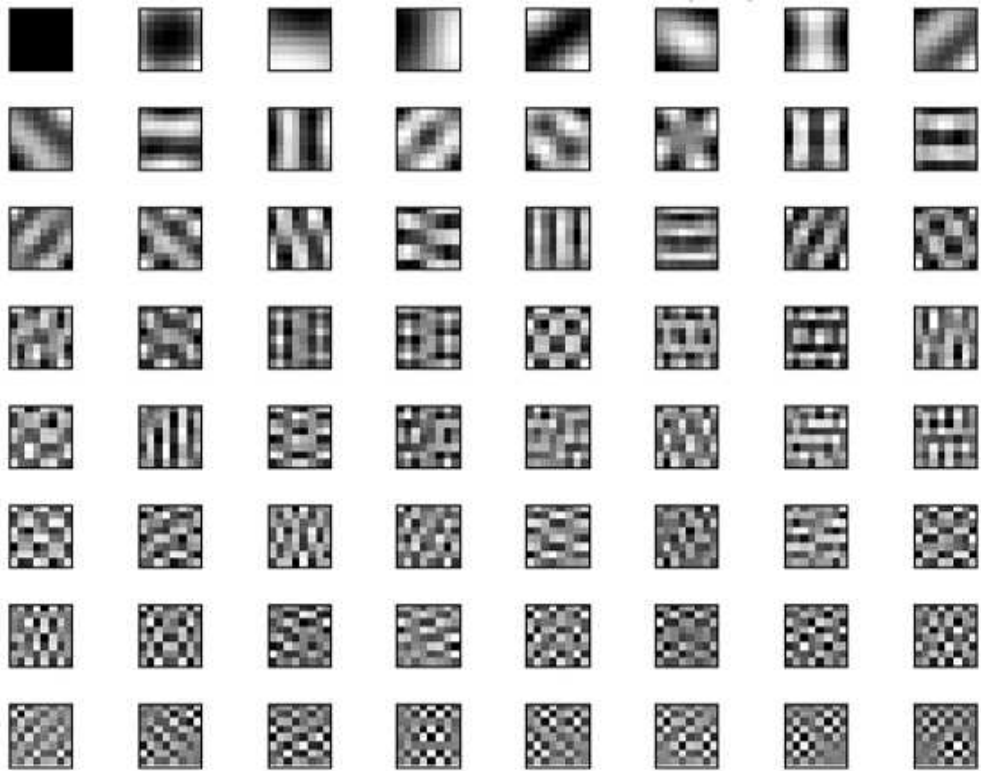}}
\caption{Visualization of basis functions of the one-stage Saab transform for
$8\times8$ residual blocks with (a) the intra horizontal mode and (b)
the intra planar mode.}\label{fig:Saab88Basis}
\end{figure*}

\noindent
{\bf Luminance Blocks of Size $8 \times 8$.}
We show the basis functions of the one-stage Saab transform for
luminance residual blocks of size $8 \times 8$ with the intra horizontal
mode and the intra planar mode in Figs. \ref{fig:Saab88Basis} (a) and
(b), respectively. When a block is predicted to be in the horizontal
mode, it tends to have textures along the horizontal direction.  Thus,
the first 8 AC basis functions in Fig. \ref{fig:Saab88Basis}(a) all have
horizontal patterns so as to express the signal better.  This is not
possible for DCT basis functions. When a block is predicted to be in the
planar mode, the basis functions can be in bowl shape, tilted planes,
saddle shape, {\em etc.}.

\begin{figure*}[!htb]
\centering
\subfigure[intra horizontal]
{\includegraphics[width=0.9\textwidth]{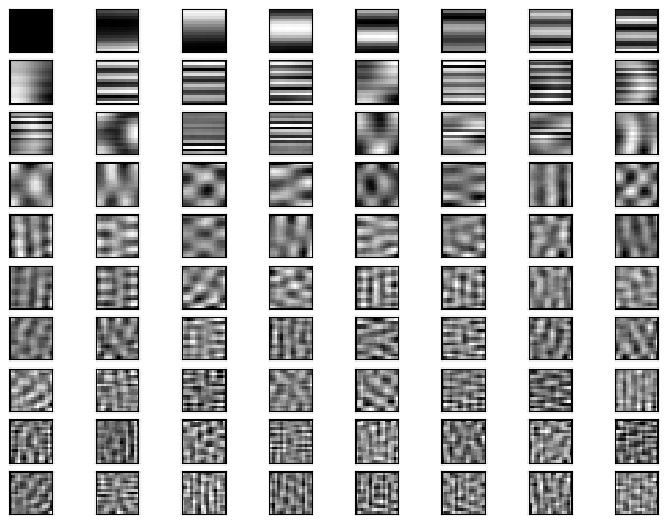}}
\subfigure[intra planar]
{\includegraphics[width=0.9\textwidth]{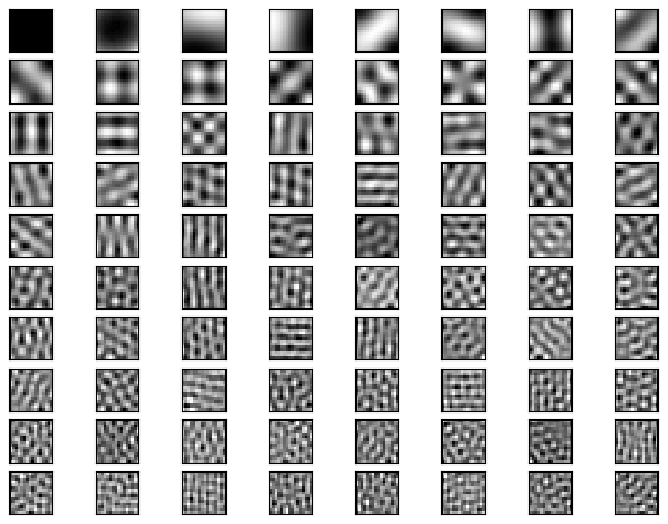}}
\caption{Visualization of the top 80 basis functions of the two-stage
Saab transform for $16\times16$ luminance residual blocks with
(a) the intra horizontal and (b) the intra planar modes.}
\label{fig:Saab16DCBasis}
\end{figure*}

\noindent
{\bf Luminance Blocks of Size $16 \times 16$.}
We show the basis functions of the two-stage Saab transform for
luminance (Y) block residuals of size $16 \times 16$ with the intra
horizontal mode and the intra planar mode in Figs.
\ref{fig:Saab16DCBasis} (a) and (b), respectively. We have observations
similar to $8 \times 8$ blocks with the intra horizontal mode and the
intra planar mode as studied above.

\section{Conclusion and Future Work}\label{sec:conclusion}

The energy compaction property of the multi-stage nonseparable
data-driven Saab transform was studied on different block residual sizes
for intra prediction in HM version 16.9 in this work. It was
demonstrated by extensive experimental results that the Saab transform
has better energy compaction capability than the widely
used DCT and PCA. Thus, we can draw the conclusion that the Saab transform
offers a highly competitive solution to the residual transform for
future image/video coding standards. Furthermore, the basis functions of
the Saab and DCT were visualized and compared. This helps explain the
advantages of the Saab transform over the DCT and the PCA.

It is desired to extend our study to block residuals of inter
prediction. Besides, we plan to incorporate the quantization and the
entropy coding modules and provide a complete picture of the
rate-distortion gain of the Saab transform over the DCT in the HEVC
video coding standard in the near future.

\section*{Acknowledgment}

This work was supported in part by Shenzhen International Collaborative
Research Project under Grant GJHZ20170314155404913 and Guangdong
International Science and Technology Collaborative Research Project
under Grant 2018A050506063. This work was also supported in part by the
National Natural Science Foundation of China (No. 61772054) and the NSFC
Key Project (No. 61632001).

\begin{thebibliography}{10}

\bibitem{Kuo2019}
C.-C.~J.~Kuo, M.~Zhang, S.~Li, J.~Duan, and Y.~Chen, ``Interpretable
convolutional neural networks via feedforward design,'' in \emph{Journal
of Visual Communication and Image Representation,} pp.346-359, 2019.

\bibitem{Su2019}
Y. Su, R. Lin, and C.-C.~J. Kuo, ``Tree-structured multi-stage principal
component analysis (TMPCA): Theory and applications,'' Expert Systems
with Applications 118 (2019): 355-364.

\bibitem{Kuo2016}
C.-C.~J.~Kuo, ``Understanding convolutional neural networks with a
mathematical model,'' in \emph{Journal of Visual Communication and Image
Representation,} vol.41, pp.406-413, 2016.

\bibitem{Kuo2017}
C.-C.~J.~Kuo, ``The CNN as a guided multilayer RECOS transform [lecture
notes],'' in \emph{IEEE Signal Processing Magazine,} vol.34, no.3,
pp.81-89, 2017.

\bibitem{Kuo2018}
C.-C.~J.~Kuo, Y.~Chen, ``On data-driven Saak transform,'' in
\emph{Journal of Visual Communication and Image Representation,} vol.50,
pp.237-246, 2018.

\bibitem{Ahmed1974}
N.~Ahmed, T.~Natarajan and K.R.~RAO, ``Discrete Cosine Transform,''
\emph{IEEE Transaction on computers,} January 1974.

\bibitem{Raj1978}
T.~Raj Natarajan and N.~Ahmed, ``Performance Evaluation for Transform
Coding Using a Nonseparable Covariance Model,'' in \emph{IEEE
Transactions on Communications,} vol. 26, no. 2, pp. 310-312, February
1978.

\bibitem{Sull2012}
G.~J.~Sullivan, J.~Ohm, W.-J.~Han, T.~Wiegand, ``Overview of the high
efficiency video coding (HEVC) standard,'' in \emph{IEEE Transactions on
Circuits and Systems for Video Technology,} 22, pp.1649¨C1668, December
2012.

\bibitem{Aston2017}
J. A.D.~Aston, D.~Pigoli and S.~Tavakoli, ``Tests for separability in
nonparametric covariance operators of random surfaces,'' in \emph{ The
Annals of Statistics,} vol.45, no.4, pp.1431-1461, 2017.

\end{thebibliography}
\end{document}